\documentclass[twocolumn,prb,amsmath,amssymb,floatfix,superscriptaddress,showpacs]{revtex4}
\usepackage{color}
\usepackage{graphicx}% Include figure files
\usepackage{dcolumn}% Align table columns on decimal point
\usepackage{bm}% bold math
\usepackage{times}
\begin{document}
% \draft command makes pacs numbers print

%\setlength{\topmargin}{0in}

\def\QAF{${\bf Q}_{\rm AF}$}

%WK
\title{Low energy magnetic excitations from the
Fe$_{1+y-z}$(Ni/Cu)$_{z}$Te$_{1-x}$Se$_{x}$ system}

\author{Zhijun~Xu}
\affiliation{Condensed Matter Physics and Materials Science
Department, Brookhaven National Laboratory, Upton, New York 11973,
USA} \affiliation{Physics Department, University of California,
Berkeley, California 94720, USA} \affiliation{Materials Science
Division, Lawrence Berkeley National Laboratory, Berkeley,
California 94720, USA}
\author{Jinsheng~Wen}
\affiliation{Center for Superconducting Physics and Materials,
National Laboratory of Solid State Microstructures and Department of
Physics, Nanjing University, Nanjing 210093, China}
\affiliation{Physics Department, University of California, Berkeley,
California 94720, USA} \affiliation{Materials Science Division,
Lawrence Berkeley National Laboratory, Berkeley, California 94720,
USA}
\author{J.~Schneeloch}
\affiliation{Condensed Matter Physics and Materials Science
Department, Brookhaven National Laboratory, Upton, New York 11973,
USA} \affiliation{Department of Physics, Stony Brook University,
Stony Brook, New York 11794, USA}
\author{A. D. Christianson}
\affiliation{Quantum Condensed Matter Division, Oak Ridge National
Laboratory, Oak Ridge, Tennessee 37831, USA}
\author{R. J. Birgeneau}
\affiliation{Physics Department, University of California, Berkeley,
California 94720, USA} \affiliation{Materials Science Division,
Lawrence Berkeley National Laboratory, Berkeley, California 94720,
USA}
\author{Genda~Gu}
\author{J.~M.~Tranquada}
\author{Guangyong~Xu}
\affiliation{Condensed Matter Physics and Materials Science
Department, Brookhaven National Laboratory, Upton, New York 11973,
USA}

\date{\today}

%WK
\begin{abstract}
We report neutron scattering measurements on low energy
($\hbar\omega \sim 5$~meV) magnetic excitations from a series of
Fe$_{1+y-z}$(Ni/Cu)$_{z}$Te$_{1-x}$Se$_{x}$ samples which belong to
the ``11'' Fe-chalcogenide family. Our results suggest a strong
correlation between the magnetic excitations near (0.5,0.5,0) and
the superconducting properties of the system. The low energy
magnetic excitations are found to gradually move away from
(0.5,0.5,0) to incommensurate positions when superconductivity is
suppressed, either by heating or chemical doping, confirming
previous observations.

\end{abstract}


\maketitle

\section{Introduction}

The role of magnetism is one of the key issues concerning the
mechanism of high temperature
superconductivity.\cite{Mazin10,pagl10,tran14} While static magnetic
order appears to compete with superconductivity, the presence of
magnetic excitations is, on the other hand, highly correlated with
the occurrence of electron pairing in high-$T_c$
cuprates\cite{Tranquada2004,Gxu2009,Rossat1991,Bourges1996,Dai2000,Fong1999,Hayden2004,Vignolle2007,Hinkov2007np}
as well as the Fe-based superconductors.\cite{pagl10,tran14,lums10r}
Direct evidence that the magnetic spins and electron pairs interact
is provided by the appearance of the ``spin
resonance"~\cite{Christianson2008,Christianson2013,Lumsden2009prl,Chis2009prl,Inosov2010nf,Qiu2009,Wen2010H}
at the superconducting phase transition---a sharp increase of the
magnetic scattering intensity at the resonance energy, $E_r$, which
is related to the size of the superconducting gap. Despite the
change of magnetic scattering intensities, the magnetic dispersion
itself, {\it i.e}. the variation of the magnetic excitation energy
with momentum, is normally not affected by superconductivity.

Recent results from the FeTe$_{1-x}$Se$_x$ system (the ``11''
system) show a surprising exception to such
behavior.\cite{zxu2012,tsyr12}  Within the superconducting phase,
low energy magnetic excitations near the in-plane wave-vector ${\bf
Q}_{\rm AF}=(0.5,0.5)$ (using the two-Fe unit cell) tend to disperse
outwards along the transverse direction with increasing energy, and
form a U-shaped dispersion, with the bottom of the dispersion, at
$E\approx E_r$, located at \QAF. When the system is heated to
temperatures well above the superconducting transition, $T_c$, this
dispersion changes to two columns, where the low energy magnetic
excitations move away from \QAF.  The incommensurate low energy
magnetic excitations are also observed for non-superconducting
compositions.\cite{zxu2012,tsyr12,Lumsden2010nf,Babkevich2010} These
results suggest an unusual connection between the locations of the
low-energy magnetic excitations in reciprocal space and the
superconducting properties of the materials.

In this paper, we report systematic studies of the low-energy
magnetic excitations in a series of single crystal samples of the
"11" system. The samples studied are listed in Table~\ref{tab:1}.
These include samples of Fe$_{1+y}$Te$_{1-x}$Se$_x$, which are
labeled with the percentage of Se and a prefix of SC, for
superconducting (with $y=0$), or NSC, for nonsuperconducting, due to
excess Fe.  Samples with Ni or Cu substitution are labeled by the
type and percentage of dopant (such as Ni02 for 2\%\ Ni
substitution); these include both superconducting and
nonsuperconducting samples.
% chosen include superconducting (SC) and
%non-superconducting (NSC) samples with A-site magnetic (Ni/Cu)
%doping; samples with/without excess Fe on the A-site; and samples
%with different B-site Se compositions.

Our results clearly show that at low temperature, the low energy
($\hbar\omega \sim 5$~meV) magnetic excitations in superconducting
samples are commensurate with \QAF, while in nonsuperconducting
samples they are split incommensurately about \QAF, as indicated
schematically in Fig.~\ref{fig:1}(b). For the nonsuperconducting
samples, there is very little change in the low-energy spectra for
temperatures between 4~K and 100~K.  In contrast, the excitations in
the superconducting samples inevitably crossover from commensurate
to incommensurate at a temperature $T^*$ well above $T_c$. The
incommensurability $\delta$ found in all samples at 100~K shows
remarkably little variation with chemical composition.  The spectral
weight of the low-energy magnetic excitations has little temperature
dependence in the normal state and also does not change much with
chemical composition. The crossover temperature $T^{*}$ varies
approximately linearly with $T_c$, further confirming its connection
to superconductivity.

\section{Experimental Details}

\begin{table}
\caption{List of the Fe$_{1+y-z}$(Ni/Cu)$_{z}$Se$_x$Te$_{1-x}$
samples used in our measurements, with their nominal composition,
superconducting transition temperature ($T_{c}$), crossover
temperature ($T^{*}$), and incommensurability $\delta$ at 100~K. }
\begin{ruledtabular}
\begin{tabular}{cccccc}
  % after \\: \hline or \cline{col1-col2} \cline{col3-col4} ...
  Sample & Compound & $T_{c}$ & $T^{*}$ & $\delta$  \\
         &          &   (K)   &   (K) & (r.l.u.)    \\
  \hline
   SC30 & FeTe$_{0.7}$Se$_{0.3}$ & 14 & 25 & 0.184  \\
   SC50 & FeTe$_{0.5}$Se$_{0.5}$ & 15 & 55 & 0.176  \\
   SC70 & FeTe$_{0.3}$Se$_{0.7}$ & 14 & 50 & 0.183  \\
   NSC45 & Fe$_{1.08}$Te$_{0.55}$Se$_{0.45}$ & --  & -- & 0.157  \\
   Ni02 & Ni$_{0.02}$Fe$_{0.97}$Te$_{0.55}$Se$_{0.45}$ & 12  & 35 & 0.200  \\
   Ni04 & Ni$_{0.04}$Fe$_{0.95}$Te$_{0.55}$Se$_{0.45}$ & 8  & 30 & 0.210  \\
   Ni10 & Ni$_{0.1}$Fe$_{0.9}$Te$_{0.55}$Se$_{0.45}$ & -- & -- & 0.181 \\
   Cu10 & Cu$_{0.1}$Fe$_{0.9}$Te$_{0.5}$Se$_{0.5}$ & --  & -- & 0.161    \\
\end{tabular}
\end{ruledtabular}
\label{tab:1}
\end{table}

\begin{figure}[t]
\includegraphics[width=0.9\linewidth]{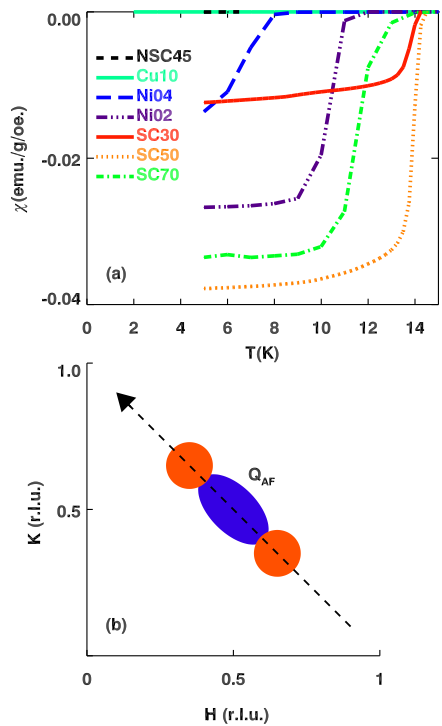}
\caption{(Color online) (a) ZFC magnetization measurements by SQUID
with a 5~Oe field perpendicular to the $a$-$b$ plane for all
samples: SC30 (red solid line), SC50 (orange dotted), SC70 (green
dot-dashed), Ni04 (blue long-dashed), Ni02 (purple 3-dot dashed),
Cu10 (teal solid) and NSC45 (black short dashed). (b) A schematic
diagram of neutron scattering measurements described in this paper.
The scattering plane corresponds to (HK0), and our constant-energy
scans are always performed along the transverse direction across
$\textbf{Q}_{\rm AF}=(0.5,0.5,0)$. The (blue) ellipse and two
(orange) circle regions are the locations of low energy magnetic
excitations measured in the superconducting and nonsuperconducting
samples, respectively.} \label{fig:1}
\end{figure}

The single-crystal samples used in this experiment were grown by a
unidirectional solidification method~\cite{JWen2011} at Brookhaven
National Laboratory. Their nominal compositions and superconducting
properties are listed in Table~\ref{tab:1}.  The bulk
susceptibilities, measured with a superconducting quantum
interference device (SQUID) magnetometer, are shown in
Fig.~\ref{fig:1}(a). Neutron scattering experiments were carried
out on the triple-axis spectrometer HB-3 located at the High Flux
Isotope Reactor (HFIR) at Oak Ridge National Laboratory (ORNL). We
used beam collimations of $48'$-$80'$-S-$80'$-$120'$ (S = sample)
with fixed final energy of 14.7~meV and a pyrolytic graphite filter
after the sample.  The inelastic scattering measurements have been
performed in the $(HK0)$ scattering plane, along the
transverse direction through \QAF, as indicated in Fig.~\ref{fig:1}(b). The lattice
constants for these sample are $a = b \approx 3.8$~\AA, and $c \approx
6.1$~\AA, using a unit cell containing two Fe atoms. The data are
described in reciprocal lattice units (r.l.u.) of $(a^*, b^*, c^*) =
(2\pi/a, 2\pi/b, 2\pi/c)$. All data have been normalized into
absolute units of $\mu_{B}^{2}{\rm eV}^{-1}$/Fe based on measurements of
incoherent elastic scattering from the samples.~\cite{GXu2013normal}
No static order around (0.5, 0, 0.5) was found in any of the these
samples, except for SC30 and NSC45.~\cite{zxu2010fetese1}

\begin{figure}[t]
\hskip20pt\includegraphics[width=0.9\linewidth]{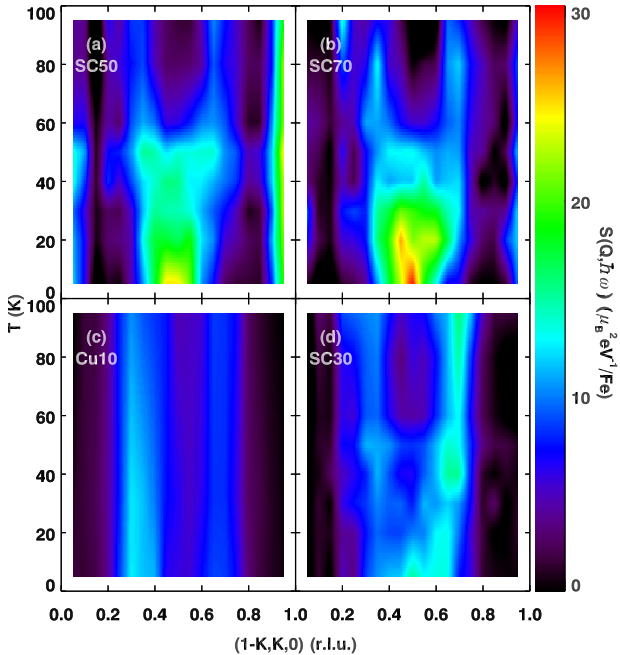}
\caption{(Color online) Thermal evolution of the magnetic scattering
at $\hbar\omega$=5~meV. The data are measured by scans through
\QAF\ along the transverse direction. Contour intensity
maps are plotted in temperature--wave-vector space for different
samples: (a) SC50, (b) SC70, (c) Cu10 and (d) SC30. The data have
been smoothed.  For (a), (b), and (d), data were measured at 5, 20, 30, 40, 50, 60, 80,100 K; for (c), measurements were at 5, 25, and 100 K.} \label{fig:2}
\end{figure}

\section{Results}

Previous work\cite{zxu2012,JWen2011} has indicated that
the low-energy spin excitations are mainly distributed along the
transverse direction about \QAF\ and that the major changes occur around the
resonance energy.  Hence, we chose to focus on constant-energy  scans
along the path shown in Fig.~\ref{fig:1} (b).
The temperature evolutions of the magnetic excitations at
$\hbar\omega=5$~meV for four samples are plotted in Fig.~\ref{fig:2}.  For
the bulk superconducting samples SC50 and SC70, shown in
Fig.~\ref{fig:2} (a)-(b), the results are similar to those from the
Ni04 sample presented in Ref.~\onlinecite{zxu2012}. The magnetic
excitation peaks clearly change from incommensurate to commensurate
upon cooling. Since the change is continuous in a broad temperature
range, it is hard to uniquely determine a crossover temperature.
We define the crossover temperature $T^*$ as the midpoint
temperature between the lowest temperature where the spectrum
clearly consists of two separated peaks, and the highest temperature
where the spectrum clearly consists of one single peak. For
nonsuperconducting sample Cu10, the results are shown in
Fig.~\ref{fig:2} (c). Similar results are obtained from NSC45 and
Ni10, where the incommensurate magnetic excitations show very little change for temperatures up to 100~K.~\cite{zxu2012} In the
case of the SC30 sample, the results are slightly more complicated
[see Fig.~\ref{fig:2} (d)]. Here, even at base temperature, the
intensity profile already shows signs of extra peaks away from
\QAF, in addition to the central peak. $T^*$ is also relatively
low despite the fact that $T_c = 14$~K is similar to the SC50 and SC70 samples. Our previous work~\cite{zxu2010fetese1}
suggests that a mixture of superconducting and nonsuperconducting phases may exist in this sample;
such phase separation has also been suggested by other
groups~\cite{He2011}. The temperature evolution can be understood
based on considering contributions from the coexisting superconducting and nonsuperconducting
regions at low temperature; all regions become nonsuperconducting above $T_c$.

\begin{figure}[t]
\hskip20pt\includegraphics[width=0.9\linewidth]{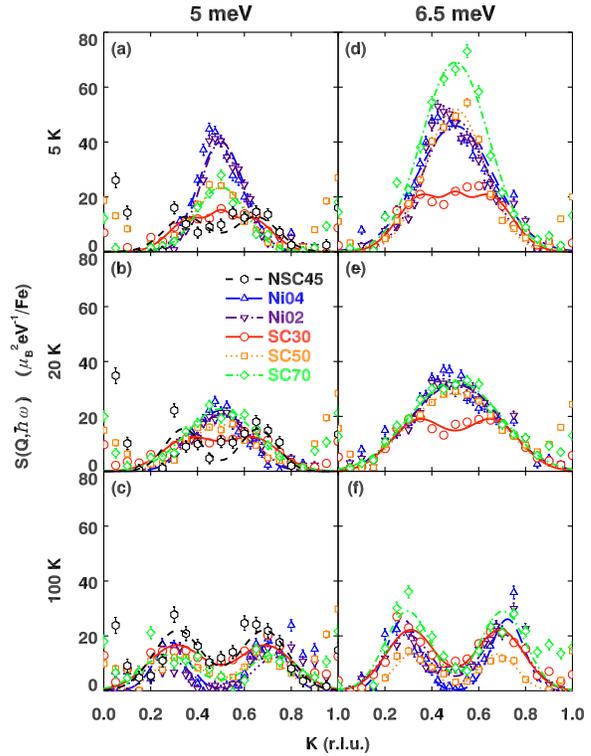}
\caption{(Color online) Constant energy scans through \QAF\ along
$(1-K,K,0)$ for different samples: SC30 (red circles), SC50 (orange
squares), SC70 (green diamonds), Ni04 (blue up triangles), Ni02
(purple down triangles) and NSC45 (black hexagons) at different
$\hbar\omega$: (a-c) 5~meV, (d-f) 6.5~meV, and different temperature
(a,d) 5~K, (b,e) 20--25~K and (c,f) 100~K. A flat fitted background
has been subtracted from all data sets. The lines are based on the
fits described in the text. The error bars represent the square root
of the number of counts.  The data for Ni02 and Ni04 are from
Ref.~\onlinecite{zxu2012}} \label{fig:3}
\end{figure}

In Fig.~\ref{fig:3}, we show constant-energy scans, for select temperatures,
performed at 5~meV and 6.5~meV. At $T=5$~K, the data for the strongly
superconducting samples show clearly commensurate single peaks; the lines through these data sets correspond to a fit by a Gaussian function. In contrast, the data
for the NSC45 sample clearly shows
incommensurate peaks, which were fit by a pair of symmetric Gaussian
functions. The data from the SC30 sample, as discussed above, were
fit by a central Gaussian function representing the contribution from
the superconducting phase, plus  a pair of symmetric Gaussian functions away from
\QAF, representing contribution from the nonsuperconducting phase.
When the superconducting samples are heated to 20~K, just above $T_c$, the
extra intensity due to the spin resonance disappears, but intensity
profiles still remain commensurate. %When the
%superconducting samples are heated up to much higher temperature at
This situation clearly changes on warming to 100~K, where the signal is split into two symmetric
incommensurate peaks. For all samples, the incommensurability of the peaks, as well as their intensities, at 100~K are remarkably similar.
%(peak positions) as well as the intensities of the data from
%non-superconducting sample show little variation for temperature
%from 5~K to 100~K. The incommensurability of all ``11" samples, no
%matter superconducting or non-superconducting, are similar.

\begin{figure}[t]
\includegraphics[width=0.9\linewidth]{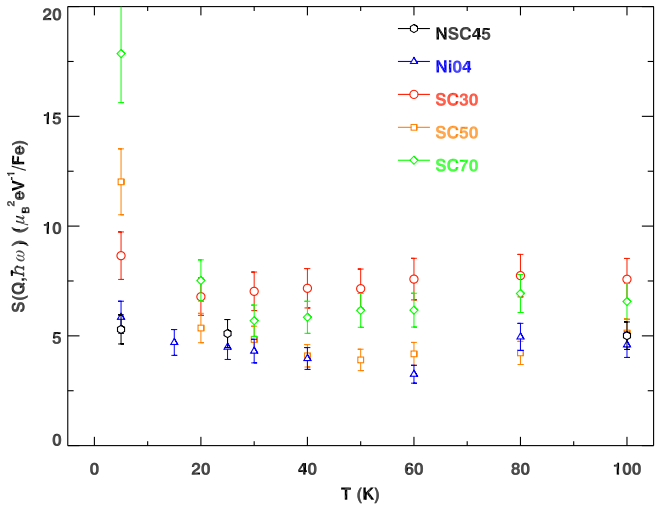}
\caption{(Color online) $\mathbf{Q}$-integrated (integrated only in
one-dimension, along the transverse direction) magnetic scattering
intensity at $\hbar\omega=6.5$~meV, obtained based on the fit
described in the text, vs. temperature for different samples: SC30
(red circles), SC50 (orange squares), SC70 (green diamonds), Ni04
(blue up triangles) and NSC45 (black hexagons).} \label{fig:4}
\end{figure}

\begin{figure}[t]
\includegraphics[width=0.9\linewidth]{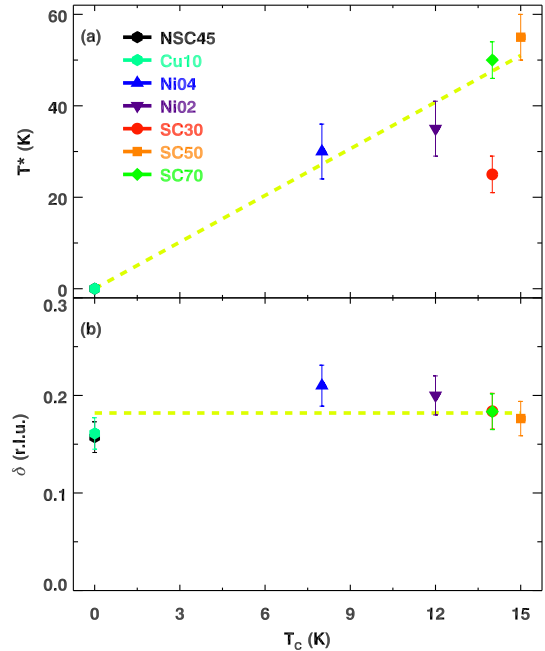}
\caption{(Color online) Summary of the fitting parameters of all
samples. (a) The cross-over temperature $T^{*}$ vs. $T_{c}$ and (b)
The incommensurability $\delta$ at 100~K vs. $T_{c}$.  SC30 (red
circles), SC50 (orange squares), SC70 (green diamonds), Ni04 (blue
up triangles), Ni02 (purple down triangles), Cu10 (teal hexagon) and
NSC45 (black hexagons). The dashed lines are guides to the eye.}
\label{fig:5}
\end{figure}

The integrated intensities of the fitted peaks are plotted as a
function of temperature in Fig.~\ref{fig:4}.   Regardless of the
sample character, the integrated intensity in the normal state shows
little temperature dependence and the major changes occur around the
superconducting transition when the spin-resonance appears. This
insensitivity of low energy spectral weight to temperature is
consistent with previous reports.\cite{zxu2011}  For the
superconducting samples, it is interesting to note that there is
little change in integrated intensity on passing through $T^*$.
%An interesting fact is that in the superconducting
%samples, the linear integrated intensity also remains virtually the
%same when the sample goes from the precursor phase ($T_c < T < T^*$)
%to the high temperature ($T>T^*$) phase with incommensurate magnetic
%excitations.
Whether this indicates a real conservation of low
energy spectral weight or is simply a coincidence cannot be resolved from these measurements,
%we cannot provide a definitive answer at this time because of the lack of a
as we need to consider the full two-dimensional intensity map. To
properly evaluate and interpret the thermal evolution of the
magnetic correlations, we will need to map them throughout the (hk0)
zone, an effort that we have just begun.

Plotting the crossover temperature $T^*$ versus $T_c$ in
%The plot of the cross-over temperature in
Fig.~\ref{fig:5}(a), we find a linear correlation between these
quantities.
% indicates that for most of our samples, there
%appears to be a direct correlation between $T_c$ and $T^*$. Samples
%with higher superconducting temperatures appear to have a higher
%cross-over temperature (from the incommensurate state to the
%commensurate state).
The only exception is the SC30 sample, which is likely due to the
complication in determining $T^*$ in this mixed-phase sample. The
incommensurability $\delta$ at 100~K is plotted in Fig.~\ref{fig:5}
(b). It is essentially independent of the superconducting
properties.

\section{Summary}

Overall, our results clearly suggest that the low energy magnetic
excitations in the ``11'' system are strongly correlated with the SC
properties. When superconductivity is destroyed with either heating
or chemical doping, the magnetic excitations move away from \QAF, becoming
incommensurate. In the normal state, the spectral weight (based on our
absolute intensity measurements) and incommensurability measured at 100~K are insensitive to the low-temperature properties,
which suggests that the incommensurates phases induced by heating or chemical
doping are qualitatively similar as far as the low energy spin
dynamics is concerned.  On the other hand, the magnetic excitations from the superconducting
phase are distinct, occurring at an entirely
different location in reciprocal space than those in the nonsuperconducting
phases. This is quite intriguing, if one considers the similarities
in the electronic structures across a range of ``11'' compounds.
ARPES measurements on various ``11'' compounds, both superconducting and nonsuperconducting,
with different Se concentrations or excess Fe,\cite{Tamai2010,Xia2009,Chen2010,Yang2013} show that the band structure near the Fermi surface is qualitatively similar
across a large doping range. The shape of the Fermi surface is
relatively invariant with Se concentration, with hole pockets near the
$\Gamma$-point, and electron pockets near the
$M$-point.~\cite{Tamai2010,Xia2009,Chen2010} No significant
change in the shape of the Fermi surface or band structure has been
reported in the temperature range of our measurements for samples without static magnetic order. The change of
low energy magnetic excitations across different samples or
different phases in the same sample, apparently are not associated with
any change of Fermi surface nesting conditions. Our results
therefore provide yet another piece of evidence that the
magnetic excitations in the ``11'' compounds cannot be simply
explained by Fermi surface topology, and contributions from both localized and itinerant
electrons have to be considered as suggested by previous
experimental and theoretical
work.~\cite{zxu2011,Yin2010,SuPeng2009,Wangm2013,dai12}

\acknowledgments

The work at Brookhaven National Laboratory and Lawrence Berkeley National Laboratory was supported by the Office of Basic Energy Sciences (BES), Division of Materials Science and Engineering, U.S. Department of Energy (DOE), under Contract Nos.\  DE-AC02-98CH10886 and DE-AC02-05CH1123, respectively. Research at Oak Ridge National Laboratory's High Flux Isotope Reactor was sponsored by the Division of Scientific User Facilities, BES, DOE.

%\bibliography{fetese,extra}

\end{document}